\documentclass[fleqn,usenatbib]{mnras}

\usepackage{newtxtext,newtxmath}

\usepackage[T1]{fontenc}

\DeclareRobustCommand{\VAN}[3]{#2}
\let\VANthebibliography\thebibliography
\def\thebibliography{\DeclareRobustCommand{\VAN}[3]{##3}\VANthebibliography}

\usepackage{graphicx}	
\usepackage{amsmath}	
\graphicspath{{./figures/}}

\title[ECSNe]{Hydrodynamic Simulations of Electron-capture Supernovae: Progenitor and Dimension Dependence}
	
\author[S. Zha et al.]{
	Shuai Zha,$^{1}$\thanks{E-mail:szha.astrop@gmail.com}
	Evan P. O'Connor,$^{2}$
	Sean M. Couch,$^{3,4,5}$
	Shing-Chi Leung,$^{6}$
	and Ken'ichi Nomoto,$^{7}$
	\\
	$^{1}$Tsung-Dao Lee Institute, Shanghai Jiao Tong University, Shanghai 200240, People's Republic of China \\
	$^{2}$The Oskar Klein Centre, Department of Astronomy, Stockholm University, AlbaNova, SE-106 91 Stockholm, Sweden \\
	$^{3}$Department of Physics and Astronomy, Michigan State University, East Lansing, Michigan 48824, USA \\
	$^{4}$Department of Computational Mathematics, Science, and Engineering, Michigan State University, East Lansing, Michigan 48824, USA \\
	$^{5}$Facility for Rare Isotope Beams, Michigan State University, East Lansing, MI 48824, USA \\
	$^{6}$TAPIR, Mailcode 350-17, California Institute of Technology, Pasadena, CA 91125, USA \\
	$^{7}$Kavli Institute for the Physics and Mathematics of the Universe (WPI), The University of Tokyo Institutes for Advanced Study, \\ The University of Tokyo, Kashiwa, Chiba 277-8583, Japan
	}

\date{Accepted 09 April 2022. Received 20 March 2022; in original form 23 December 2021}

\pubyear{2021}

\begin{document}
\label{firstpage}
\pagerange{\pageref{firstpage}--\pageref{lastpage}}
\maketitle

	
\begin{abstract}
We present neutrino-transport hydrodynamic simulations of electron-capture supernovae (ECSNe) in \texttt{FLASH} with new two-dimensional (2D) collapsing progenitor models. These progenitor models feature the 2D modelling of oxygen-flame propagation until the onset of core collapse. We perform axisymmetric simulations with 6 progenitor models that, at the time of collapse, span a range of propagating flame front radii. For comparison, we also perform a simulation with the same setup using the canonical, spherically-symmetrical progenitor model n8.8. We found that the variations in the progenitor models inherited from simulations of stellar evolution and flame propagation do not significantly alter the global properties of the neutrino-driven ECSN explosion, such as the explosion energy ($\sim1.36$--$1.48\times10^{50}$~erg) and the mass ($\sim0.017$--$0.018M_\odot$) and composition of the ejecta. Due to aspherical perturbations induced by the 2D flame, the ejecta contains a small amount ($\lesssim1.8\times10^{-3}~M_\odot$) of low-$Y_e$ ($0.35<Y_e<0.4$) component. The baryonic mass of the protoneutron star is $\sim1.34~M_\odot$ ($\sim1.357~M_\odot$) with the new (n8.8) progenitor models when simulations end at $\sim400$~ms and the discrepancy is due to updated weak-interaction rates in the progenitor evolutionary simulations. Our results reflect the nature of ECSN progenitors containing a strongly degenerate ONeMg core and suggest a standardized ECSN explosion initialized by ONeMg core collapse. Moreover, we carry out a rudimentary three-dimensional simulation and find that the explosion properties are fairly compatible with the 2D counterpart. Our paper facilitates a more thorough understanding of ECSN explosions following the ONeMg core collapse, though more three-dimensional simulations are still needed. 
\end{abstract}



\section{Introduction}
Electron-capture supernovae (ECSNe) have been proposed as the complementary channel for producing neutron stars \citep[NS; see][]{1980PASJ...32..303M,1984ApJ...277..791N}, other than the canonical channel, i.e. core-collapse supernova \citep[CCSN; see the recent review by][]{2021Natur.589...29B}. This additional channel can explain some bi-modality in the properties of observed NSs, such as the spin-orbit period \citep{2011Natur.479..372K} and mass distribution \citep{2010ApJ...719..722S,2019ApJ...876...18F}. Moreover, NSs born in ECSNe are believed to obtain a small kick velocity, $\lesssim100$~km/s \citep{2004ApJ...612.1044P,2018ApJ...865...61G,2020MNRAS.496.2039S}, so the ECSN channel can be rather important for making compact binaries with a NS component \citep{2018MNRAS.481.1908K}. 

Progenitors of ECSNe have a zero-age-main-sequence (ZAMS) mass of $\sim$8--10$~M_\odot$, lying just below that of CCSN progenitors \citep{1988PhR...163...13N,2002RvMP...74.1015W}. Therefore, ECSNe can be abundant due to the bottom-heavy initial mass function of stars \citep{1955ApJ...121..161S} and thus contribute to a significant portion of supernova-like transients. Depending on the mass loss during the last evolutionary stage of their progenitors, ECSNe may lead to normal to low-luminosity type IIp supernovae \citep{2009MNRAS.398.1041B,2013ApJ...771L..12T,2021MNRAS.503..797K,2021NatAs...5..903H}, Type IIn supernovae \citep{2014A&A...569A..57M}, fast evolving luminous transients \citep{2019ApJ...881...35T}, and intermediate-luminosity red transients \citep{2021arXiv210805087C}.

Unlike a CCSN, an ECSN progenitor contains a degenerate oxygen-neon-magnesium (ONeMg) core at its final evolutionary stage \citep{2017hsn..book..483N}. An oxygen flame develops due to the heating by electron captures on $^{24}$Mg and $^{20}$Ne, and during its outward propagation as a deflagration wave, electron captures on the nuclear-statistical-equilibrium (NSE) ash behind the flame front lead to the collapse of the ONeMg core. However, from the beginning of ECSN studies, it has been debated in theory that the oxygen flame may as well lead to a (partially) thermonuclear explosion of ONeMg core, similar to that in a Type Ia supernova \citep{1991ApJ...367L..19N,1992ApJ...396..649T,2016A&A...593A..72J,2019PASA...36....6L,2019ApJ...886...22Z,2019PhRvL.123z2701K}. 

The collapse-explosion puzzle of ECSNe cannot be solved solely from the theoretical side, due to the subtleties involved in simulations, such as the weak-reaction rates \citep{2019ApJ...881...64S,2021arXiv210402614S,2021RPPh...84f6301L}, convective URCA process \citep{1973ApL....15..147P,2015MNRAS.447.2696D,2017ApJ...851..105S}, convective mixing process driven by oxygen burning \citep{1987ApJ...322..206N,2019ApJ...886...22Z,2019ApJ...871..153T}, and physics of the oxygen flame \citep{2020ApJ...889...34L,2020ApJ...891....5S}. Each scenario, i.e. collapse or thermonuclear explosion, needs to be examined carefully with sophisticated simulations, which make solid predictions to confront the existing and upcoming observations. 

For the collapse scenario of ECSN, hydrodynamic simulations of the collapse and explosion phases \citep{2006A&A...450..345K,2008A&A...485..199J,2010A&A...517A..80F,2017ApJ...850...43R,2020MNRAS.496.2039S} have been using the spherically symmetrical progenitor model n8.8 provided in \cite{1984ApJ...277..791N,1987ApJ...322..206N}. This canonical ECSN progenitor model has led to substantial knowledge on the explosion properties of ECSNe. However, important updates have been made to the weak-interaction rates \citep{2013PhRvC..88a5806T,2014PhRvC..89d5806M,2016ApJ...817..163S,2019PhRvL.123z2701K,2019ApJ...881...64S} and the numerical schemes for the flame propagation, e.g. more than 1D modelling \citep{2016A&A...593A..72J,2020ApJ...889...34L}. In this work, we perform a suite of 2D axisymmetric neutrino-transport simulations of ECSNe with the updated progenitor models possessing a collapsing ONeMg core provided in \cite{2019ApJ...886...22Z}. These progenitor models feature the updated weak-interaction rates and 2D simulations for the propagation of the oxygen flame. We focus on how the uncertainties in the progenitor evolution transforms into the explosion properties of ECSNe. Moreover, we carry out a rudimentary 3D octant simulation to examine the 3D effects. Our results show that unlike CCSNe whose progenitors are more massive, the 3D ECSN simulation exhibits only small differences compared to its 2D axisymmetric counterpart. 

Our paper is organized as follows. In Section~\ref{sec:method} we describe the ECSN progenitor models and methods used in our numerical simulations. Results of progenitor and dimension dependence are presented in Section~\ref{sec:result} and conclusions are given in Section~\ref{sec:conclu}.

\section{Methods} \label{sec:method}
\subsection{Progenitor models} \label{subsec:prog}
We use the collapsing ECSN progenitor models of \cite{2019ApJ...886...22Z}. Here we briefly describe their evolution prior to collapse for the sake of completeness, and refer readers to \cite{2019ApJ...886...22Z} and \cite{2020ApJ...889...34L} for details.

A solar-metallicity star with a ZAMS mass of $8.4~M_\odot$ was evolved with the \texttt{MESA} code \citep{Paxton2011,Paxton2013,Paxton2015,Paxton2018,Paxton2019} from the main-sequence phase to the formation of a degenerate ONeMg core. The simulation of the whole star was terminated when its helium (He) layer becomes thin ($\simeq0.2 M_\odot$). Afterwards the hydrogen (H) envelope and He layer were removed and the ONeMg core was evolved with a constant mass accretion rate, $10^{-6}$ or $10^{-7}$~$M_\odot$~yr$^{-1}$, to mimic the H-He thermal pulses. During the growth of the ONeMg core, two extreme criteria, i.e. Ledoux and Schwarzschild \citep{1987ApJ...318..307M,2012sse..book.....K}, were used to explore the uncertain semiconvection driven by the heating due to electron captures on $^{24}$Mg and $^{20}$Ne. The \texttt{MESA} calculation was terminated at the ignition of oxygen burning, when its energy release exceeds that of thermal neutrino losses. The ONeMg core has a central density of $\sim10^{10}$~g~cm$^{-3}$ at this moment.

The oxygen burning will eventually turn into nuclear runaway and initiation of a deflagration wave that propagates the oxygen-flame front. On the other hand, oxygen burning can drive convection to allow energy being transported away from the burning region and thus defer the nuclear runaway to a higher central density. Two approaches were used to follow the evolution after the oxygen ignition. Firstly, the nuclear runaway was assumed to happen immediately after the oxygen ignition. Secondly, to accommodate the effect of the convection driven by oxygen burning, the last \text{MESA} profile was modified with a fully mixed central region of $\sim0.14~M_\odot$, while the central density was parameterized with a larger value between $10^{10}$ to $10^{10.2}$~g~cm$^{-3}$.

The last \texttt{MESA} profile was then mapped into a 2D cylindrical grid with an initial flame sitting at the spot of oxygen burning. A 2D hydrodynamic code with the capability of flame-capturing \citep{2015MNRAS.454.1238L,2020ApJ...889...34L} was used to follow the propagation of the oxygen flame and electron captures on the NSE ash behind it. The simulations were terminated when the ONeMg core reaches a central density of $3\times10^{10}$~g~cm$^{-3}$, signaling its collapse.

\begin{figure*}
    \centering
    \includegraphics[width=0.95\textwidth]{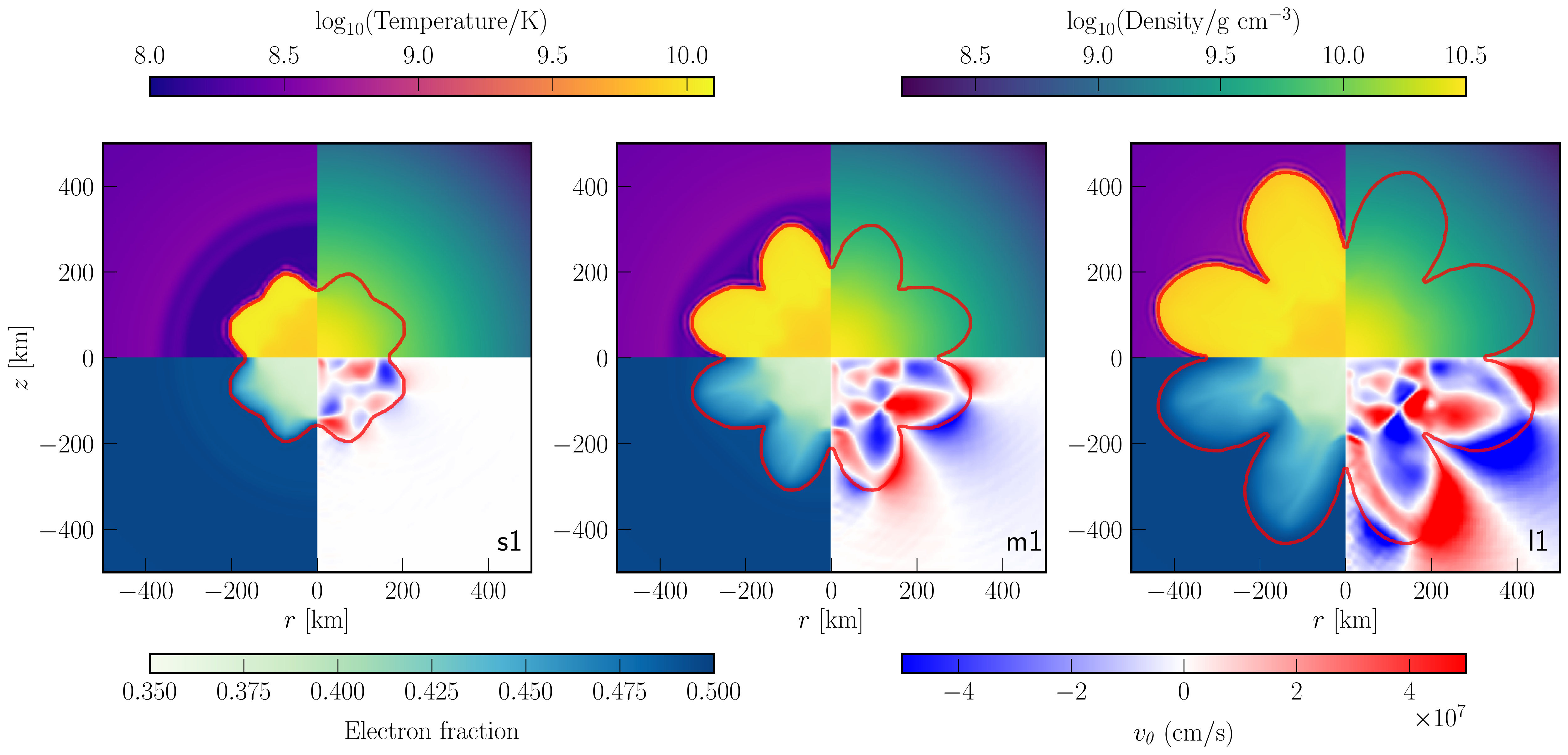}
    \caption{Profiles of density, temperature, electron fraction, and transverse velocity ($v_\theta$) for the progenitor models s1, m1 and l1. The red line in each panel marks the locus of the oxygen-flame front in the progenitor models. Note that the models are only in the first quadrant, and the variables except density are mirrored to other quadrants.}
    \label{fig:prog}
\end{figure*}

\begin{table*}
    \centering
    \begin{tabular}{ccccccccccc}
    \hline
         Model & {$M_{\rm core}$}  & {$R_{\rm core}$}  & {$v_{\rm infall}$}& {$M_{\rm NSE}$} & {$R_{\rm flame,\max}$} & {$R_{\rm flame,\min}$}  &  {$\langle R_{\rm flame}\rangle$}  & {$E_{\rm kin,\theta}$} & {Nomenclature in}  & {Mixing scheme}\\ 
    & {($M_\odot$)} & {(km)} & {($10^8$~cm~s$^{-1}$)}  & {($M_\odot$)} & {(km)}   & {(km)}   &{(km)} & ($10^{46}$erg) & {\cite{2019ApJ...886...22Z}}&  \\ \hline
         n8.8 & 1.377 & 1090 & 0.8 & 0.090 & 110 & 110 & 110 & N.A. & N.A. & N.A.\\
         s1 & 1.361 & 1190 & 1.6 & 0.270 & 219 & 158 & 198 & 7.09 & 6-1020-046-30 & L\_no\_mix \\
         s2 & 1.361 & 1180 & 1.6 & 0.250 & 208 & 166 & 191 & 3.65 & 7-1010-046-00 & L\_no\_mix\\
         m1 & 1.359 & 1210 & 2.4 & 0.539 & 338 & 212 & 308 & 41.7 & 6-0999-049-00 &  S\_$\rho$\_mix \\
         m2 & 1.358 & 1180 & 2.1 & 0.498 & 319 & 205 & 294 & 32.6 & 7-0997-049-00 & S\_$\rho$\_mix \\
         l1 & 1.358 & 1200 & 3.3 & 0.790 & 460 & 254 & 363 & 117  & 6-1000-049-00 & S\_$\rho$\_mix\\
         l2 & 1.359 & 1210 & 3.0 & 0.709 & 426 & 209 & 341 & 108  & 6-0999-049-30-LM & Ledoux mix o-burn \\ \hline
    \end{tabular}
    \caption{Properties of the progenitor models used in this work. $M_{\rm core}$ and $R_{\rm core}$ are the mass and radius of the ONeMg core. $v_{\rm infall}$ is the maximum infall velocity in the ONeMg core at collapse. $M_{\rm NSE}$ is the mass of the NSE ash enclosed by the flame front. $R_{\rm flame,\max}$, $R_{\rm flame,\min}$ and $\langle R_{\rm flame}\rangle$ are the maximum, minimum and angular-averaged radius of the flame front. $E_{\rm kin,\theta}$ is the total transverse kinetic energy of the progenitor models. The corresponding nomenclature and mixing scheme used in Section 3.2 of \citet{2019ApJ...886...22Z} are given for clarity. ``L\_no\_mix" and ``S\_$\rho$\_mix" refer to \texttt{MESA} simulations using the Ledoux and Schwarzschild criteria for the convective instability, respectively. They both assumes nuclear runaway occurs immediately after the oxygen ignition. ``Ledoux mix o-burn" uses the Ledoux criterion and considers the convection driven by the oxygen burning.}
    \label{tab:prog}
\end{table*}

\cite{2019ApJ...886...22Z} found that the configuration of the ONeMg core at the oxygen runaway determines the subsequent flame propagation. In general, a higher central density and a more mixed central region favor electron captures, accelerate core contraction, and result in a shorter distance of the flame propagation until the onset of core collapse. Here we further investigate how the propagated distance of the oxygen flame will affect properties of ECSN explosions. In Table~\ref{tab:prog} we list the progenitor models of \cite{2019ApJ...886...22Z} employed in this work. The model names indicate the size of the region enclosed by the flame front, with s, m and l corresponding to small, medium and large. Our nomenclature for the models is different from that in Section 3.2 of \cite{2019ApJ...886...22Z}, which is also given in Table~\ref{tab:prog} for clarity. We choose 2 models for each group of flame size and the mean radius of the flame front in the model `2' is smaller by $\sim10-20$~km than that in the model `1'. We diversify the progenitor models with respect to the parameters used in the stellar evolutionary calculations \citep{2019ApJ...886...22Z}. For example, in groups `s' and `m', the model `1' uses a mass accretion rate of $10^{-6}~M_\odot~{\rm yr}^{-1}$ and the model `2' uses a mass accretion rate of $10^{-7}~M_\odot~{\rm yr}^{-1}$.  These models cover the extremes of propagation of the oxygen flame and reflect the uncertainties in the progenitor evolutionary simulations. We note that the mass of the NSE ash enclosed by the flame front ranges from $18\%$ to $58\%$ of the ONeMg core. The models used here are publicly available at \url{https://doi.org/10.5281/zenodo.5748457}. We also perform simulations with the progenitor model n8.8 \citep{1984ApJ...277..791N,1987ApJ...322..206N} for comparison. One can see from Table~\ref{tab:prog} that n8.8 has the smallest mass of NSE ash and radius of the flame front.

The 2D profiles of density, temperature, electron fraction ($Y_e$), and transverse velocity ($v_\theta$) are rendered in Fig.~\ref{fig:prog} for progenitor models s1, m1 and l1. The red line in each plot marks the locus of the oxygen-flame front in the progenitor models. 

\subsection{Neutrino-transport simulations} \label{subsec:sim}
We utilize the \texttt{FLASH} code \citep{2000ApJS..131..273F} for the neutrino-transport hydrodynamic simulations that follow the collapse of the ONeMg core and the subsequent explosion. The equation of state (EOS) of the stellar matter is described by the SFHo EOS \citep{2013ApJ...774...17S} at densities above $4\times10^{6}$~g~cm$^{-3}$ and the Helmholtz EOS \citep{2000ApJS..126..501T} at densities below $2\times10^{6}$~g~cm$^{-3}$, with a linear interpolation for smooth transition in between \citep{2021ApJ...921...19W}. Neutrino transport with full velocity dependence is simulated with a two-moment scheme and the analytic ``M1" closure \citep{2011PThPh.125.1255S,2013PhRvD..87j3004C,2015ApJS..219...24O,2018ApJ...854...63O}. Neutrino-matter interactions are described by \texttt{NuLib} \citep{2015ApJS..219...24O}, with 12 energy groups from 0 to 300~MeV spaced  logarithmically. We use the same baseline prescription for neutrino-matter interactions as that of \citep{2018JPhG...45j4001O}. We note explicitly that these FLASH simulations, unlike the FLASH simulations in \citep{2018JPhG...45j4001O}, include neutrino-electron inelastic scattering.

We use a block structured adaptive-mesh-refinement grid for our simulations. In the 2D simulations, the cylindrical grid extends to $1.2\times10^4$~km and includes the first quadrant, assuming reflective symmetry at the equatorial plane as in the previous simulations of the  deflagration phase \citep{2019ApJ...886...22Z}. Our fiducial resolution is $\sim$250~m at $\lesssim40$~km to resolve the protoneutron star (PNS) and maintains an effective angular resolution of 0.6$^{\circ}$ at $\gtrsim40$~km. In the 3D octant simulation, we use a coarser resolution, $\sim$500~m at the center and $1.5^{\circ}$ outside $40$~km. To disentangle the effects of resolution and dimension, we run a 2D simulation (m1-2d-lr) with a comparable resolution as the 3D run (m1-3d-lr). The timestep is smaller in 3D (0.43~$\mu $s) than that in 2D (0.65~$\mu$s) to maintain numerical stability of the neutrino transport. The results of integrated quantities such as the explosion energy and mass of ejecta are multiplied by 2 in 2D and 8 in 3D according to the imposed symmetries. In addition, we perform two 1D simulations with the fiducial resolution using n8.8 (n8.8-1d) and spherically-averaged m1 (m1-1d) as the progenitor models.

In the simulations of the deflagration phase the computational domain only included the ONeMg core. To follow the evolution of shock after its breaking out of the core surface, we attach a convective envelope above the core with $X_{\rm H}=0.6$ and $X_{\rm He}=0.4$, where $X_i$ is the mass fraction of element $i$. The envelope has a mass of $\lesssim10^{-5}~M_\odot$ and should not affect the results of the ECSN explosion, such as the explosion energy and the properties of the ejecta \citep{2008A&A...485..199J}.

\section{Results} \label{sec:result}
In this section, we first present a systematic investigation of the progenitor dependence with 2D axisymmetric simulations. Then we compare the results of the 2D and 3D octant simulations with the same progenitor and show the impact of the 3rd dimension. We further compare the properties of the ejecta in different simulations and discuss the implications for nucleosynthesis of ECSNe. Finally we discuss the caveat that our neutrino-transport simulations lacks a nuclear network to simulate the flame and burning of oxygen and neon.

\subsection{Axisymmetric simulations and progenitor dependence} \label{subsec:hydro}
\begin{figure}
    \centering
    \includegraphics[width=0.47\textwidth]{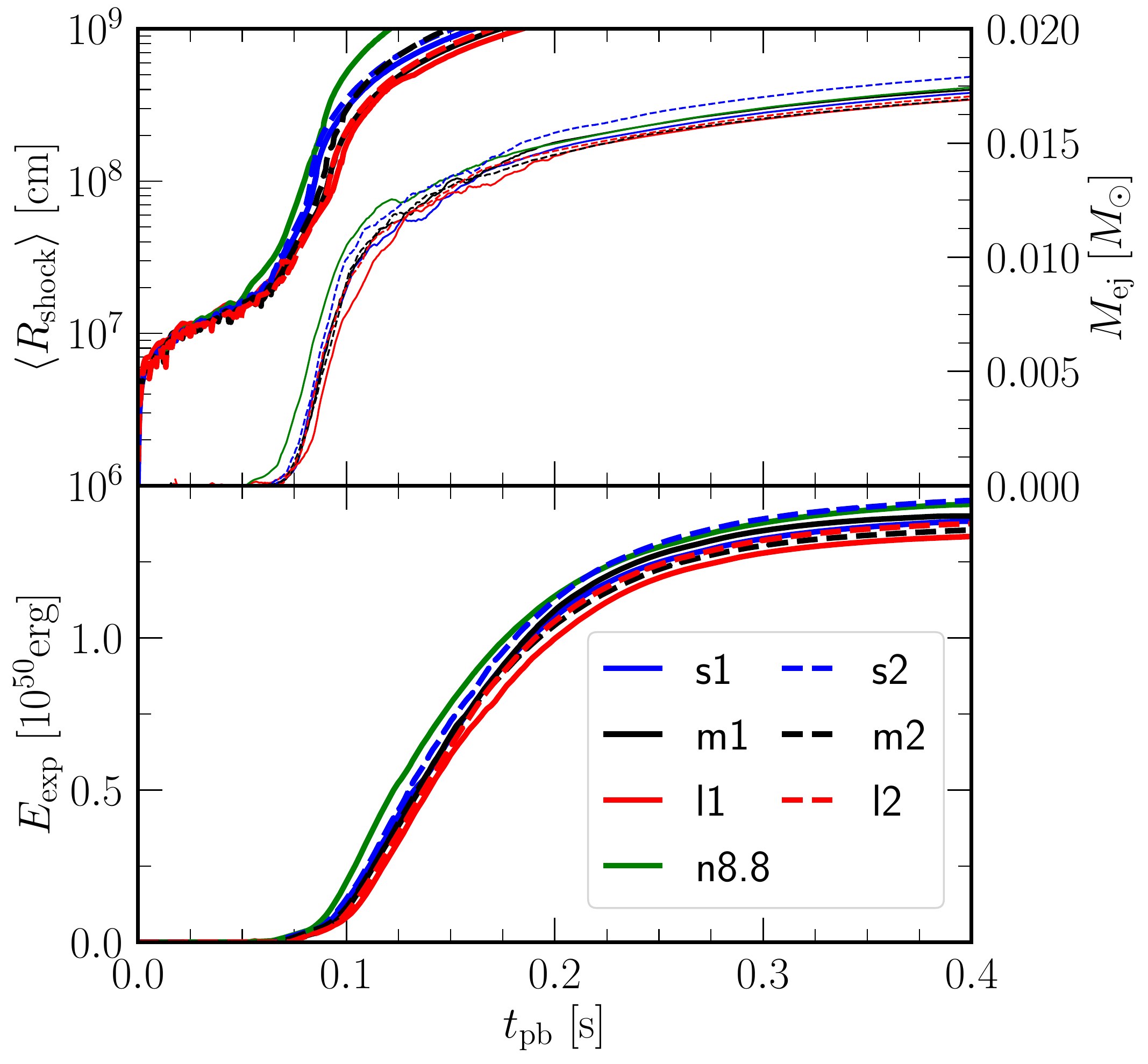}
    \caption{Mean shock radius ($\langle R_{\rm shock}\rangle$, upper panel, thick lines), mass of ejecta ($M_{\rm ej}$, upper panel, thin lines), and diagnostic explosion energy ($E_{\rm exp}$, lower panel) as a function of the postbounce time in the axisymmetric simulations.}
    \label{fig:exp}
\end{figure}

\begin{table}
    \centering
    \begin{tabular}{ ccccc}
    \hline
    {Model} & {$t_{\rm b}$}  & $E_{\rm exp}$  &  $M_{\rm ej}$ & $M_{\rm PNS}$     \\ 
    & {(ms)}  & {$(10^{51}\rm erg)$}  &  {$(10^{-2}M_\odot)$} & {$(M_\odot)$}  \\ \hline
         n8.8-1d  & 62.2 & 0.111 & 1.39 & 1.362    \\
         m1-1d  & 63.8 & 0.109 & 1.35 & 1.344    \\ \hline
         n8.8  & 62.8 & 0.146 & 1.74 & 1.357    \\
         s1    & 68.5 & 0.141 & 1.72 & 1.341    \\
         s2    & 69.7 & 0.148 & 1.79 & 1.340    \\
         m1    & 67.2 & 0.143 & 1.73 & 1.339    \\
         m2    & 66.5 & 0.139 & 1.69 & 1.339    \\
         l1    & 75.8 & 0.136 & 1.69 & 1.340    \\
         l2    & 71.1 & 0.136 & 1.70 & 1.340    \\ \hline
         m1-2d-lr & 67.2 & 0.188 & 2.25 & 1.333    \\
         m1-3d-lr & 67.1 & 0.177 & 2.06 & 1.334    \\ \hline
    \end{tabular}
    \caption{Results of the hydrodynamic simulations. n8.8-1d and m1-1d are the 1D simulations with the progenitor models n8.8 and spherically-averaged m1, respectively. s1 to l2 are the 2D axisymmetric simulations with the corresponding progenitor models. These are with the fiducial resolution. m1-2d-lr and m1-3d-lr are the low-resolution simulations with the progenitor model m1 in 2D and 3D, respectively. $t_{\rm b}$ is time of bounce. $E_{\rm exp}$, $M_{\rm ej}$ and $M_{\rm PNS}$ are the diagnostic explosion energy, mass of ejecta and baryonic mass of protoneutron star at the end of simulations, which is 400~ms after bounce except for m1-3d-lr and m1-2d-lr which are at 300~ms after bounce).}
    \label{tab:hydro}
\end{table}

Due to the high initial central density ($3\times10^{10}$~g~cm$^{-3}$) of ECSN progenitors, the collapse of ONeMg core lasts for only $\sim$60-70~ms until the formation of the protoneutron star (PNS) and the bounce shock wave. The bounce time, $t_{\rm b}$, as listed in Table~\ref{tab:hydro}, differs by 10~ms at most for 2D progenitors and shows a non-monotonic trend with the size of flame front due to the competition between electron captures and thermal effects. Model n8.8 has a central density of $4.43\times10^{10}$~g~cm$^{-3}$ and the smallest $t_{\rm b}$ of 62.8~ms. As plotted in the upper panel of Fig.~\ref{fig:exp}, the bounce shock reaches a radius of $100$~km in $\sim20$~ms, and then gradually expands until breaking out of the core surface at $\sim450$~km and $\sim80$~ms postbounce. Due to the steep density gradient at the core surface, the shock gets accelerated there and reaches 1000~km in another 20~ms. 

In the lower panel of Fig.~\ref{fig:exp} we plot the instantaneous diagnostic explosion energy, defined as 
\begin{equation}
    E_{\rm exp} = \int_{e_{\rm tot}>0,v_r>0} dV \rho e_{\rm tot}, 
\end{equation}
where $\rho$ and $v_r$ are the matter density and radial velocity of a fluid element. $e_{\rm tot}$ is the specific total energy
\begin{equation}
e_{\rm tot} = e-e_{\rm cold}+\Phi+0.5 |\Vec{v}|^2,  
\end{equation}
where $e$, $\Phi$ and $0.5 |\Vec{v}|^2$ are the specific internal, gravitational and kinetic energy. $e_{\rm cold}$ is the the specific internal energy of matter with the same density and $Y_e$ of the fluid element, but with temperature set to be the minimum value of the SFHo EOS table (0.01~MeV). Once the shock breaks out of the core surface, $E_{\rm exp}$ quickly becomes nonzero due to neutrino heating and gradually increases to $\sim1.3\times10^{50}$~erg in the following $\sim200$~ms. Afterwards $E_{\rm exp}$ increases slowly to $\sim1.4\times10^{50}$~erg at the end of simulations and differs by $0.12\times10^{50}$~erg ($\sim$9$\%$) at most among all progenitor models (cf. Table~\ref{tab:hydro}). 

The mass of the ejecta, defined as 
\begin{equation}
    M_{\rm ej} = \int_{e_{\rm tot}>0,v_r>0} dV \rho,
\end{equation}
is $\sim$0.017-0.018$~M_\odot$ at the end of the simulations (thin lines in the upper panel of Fig.~\ref{fig:exp}). We note that due to the shock breakout of the outer boundary of our computational domain at $\sim150$~ms postbounce, a small amount of matter with mass$\sim2\times10^{-4}~M_\odot$ and $E_{\rm exp}\sim3\times10^{48}$~erg has left the grid by the end of the simulations. This accounts for $\sim1\%$ of $M_{\rm ej}$ and $\sim2\%$ of $E_{\rm exp}$ which we add back in Fig.~\ref{fig:exp} and Table~\ref{tab:hydro}. As shown in Fig.~\ref{fig:exp} and Table~\ref{tab:hydro}, the explosion properties show a modest dependence on the ECSN progenitor, while 2D simulations have a larger explosion energy and ejecta mass by 25\%-30\% comparing to their 1D counterparts (n8.8-1d and m1-1d). In Section~\ref{subsec:ejecta} we will further show the detailed composition of the ejecta and the implications for ECSN nucleosynthesis.

\begin{figure*}
    \centering
    \includegraphics[width=0.49\textwidth]{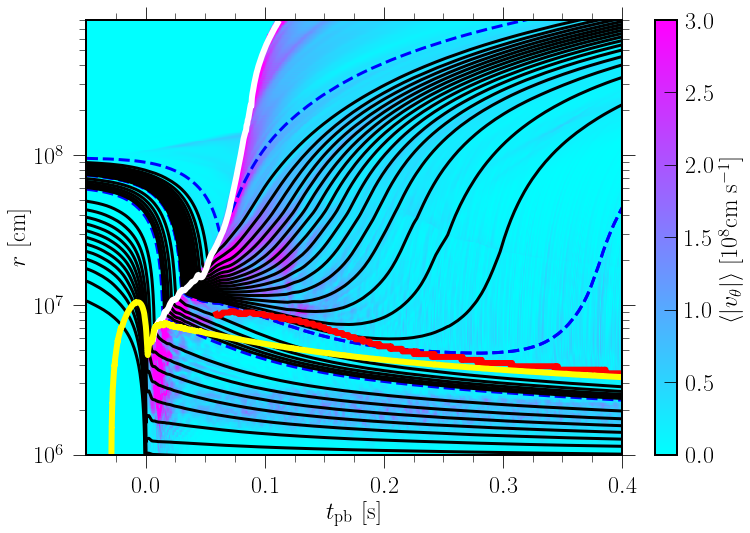}
    \includegraphics[width=0.49\textwidth]{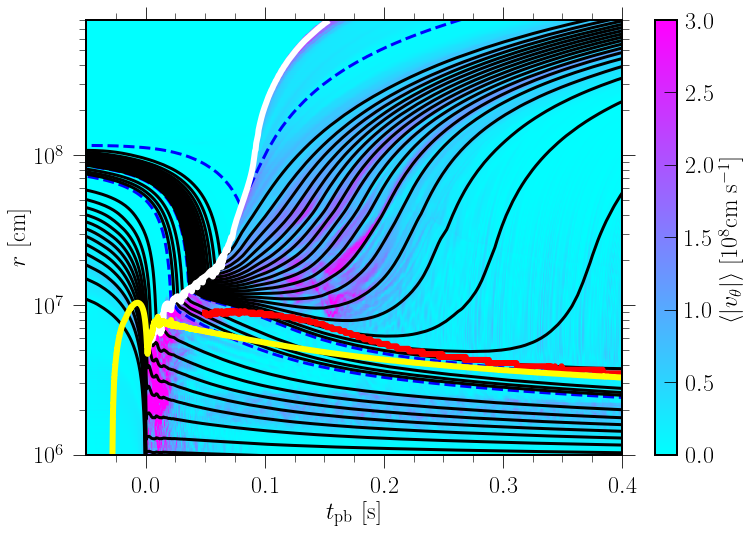}
    \caption{Colormaps depicting the magnitude of angular averaged transverse velocity $\langle |v_\theta|\rangle$ as a function of radius and the postbounce time in model n8.8 (left) and m1 (right). Black lines indicate the radii that enclose specific masses. The mass resolution in the left (right) plot is 0.1~$M_\odot$, 0.01~$M_\odot$ and 0.001~$M_\odot$ below $M=$1.3 (1.3), 1.36 (1.34) and 1.377 (1.359)~$M_\odot$, which are indicated by the blue dashed lines. The white, red and yellow lines mark the mean shock radius, mean gain radius and PNS surface; the radius of the PNS surface is defined as the mean radius with a density of $10^{11}~{\rm g~cm}^{-1}$.}
    \label{fig:massshell2d}
\end{figure*}

Recent CCSN simulations found that precollapse aspherical perturbations in the convective oxygen and silicon shells \citep{2020ApJ...901...33F,2021ApJ...921...28F,2017MNRAS.472..491M,2022MNRAS.510.4689V} are amplified when crossing the standing accretion shock and induce violent turbulent convection which can enhance the neutrino energy deposition \citep[see e.g. ][]{2013ApJ...778L...7C,2017MNRAS.472..491M,2018ApJ...865...81O,2018MNRAS.477.3091V}. Our progenitor models also possess such perturbations inside and around the oxygen-flame front (see bottom right corner of Fig.~\ref{fig:prog}). To see how this affects the postbounce turbulence, in Fig.~\ref{fig:massshell2d} we plot the angular-averaged transverse velocity $\langle |v_\theta|\rangle$ as a function of radius and the postbounce time in models n8.8 and m1. As the bounce shock is launched the postshock region gains a $\langle |v_\theta|\rangle$ up to $\sim3\times10^8$~cm~s$^{-1}$ in both models. This is the so-called prompt convection due to the negative entropy gradient behind the shock \citep{1994ApJ...433L..45B}. This prompt convection spans from $\sim10$~km to the shock and lasts for $\sim20$~ms. Model m1 shows a slightly stronger turbulence than model n8.8 and this is compared quantitatively later. After this episode, convection develops in the neutrino-heated postshock region and below the PNS surface, similar to that in the simulations of massive progenitors. Here, we define the PNS surface as the mean radius with a density of $10^{11}~{\rm g~cm}^{-3}$. The magnitudes of $\langle |v_\theta|\rangle\sim1.5\times10^{8}$~cm~s$^{-1}$ are quantitatively similar in the two models for both convective regions. We note a similar magnitude of $\langle |v_\theta|\rangle$ just below the fast expanding shock wave (white lines in Fig.~\ref{fig:massshell2d}) especially for model n8.8. The transverse kinetic energy of this fluid element is small ($\sim1\%$) comparing to that in the gain region due to its small mass ($\sim10^{-4}~M_\odot$). 

\begin{figure*}
    \centering
    \includegraphics[width=0.47\textwidth]{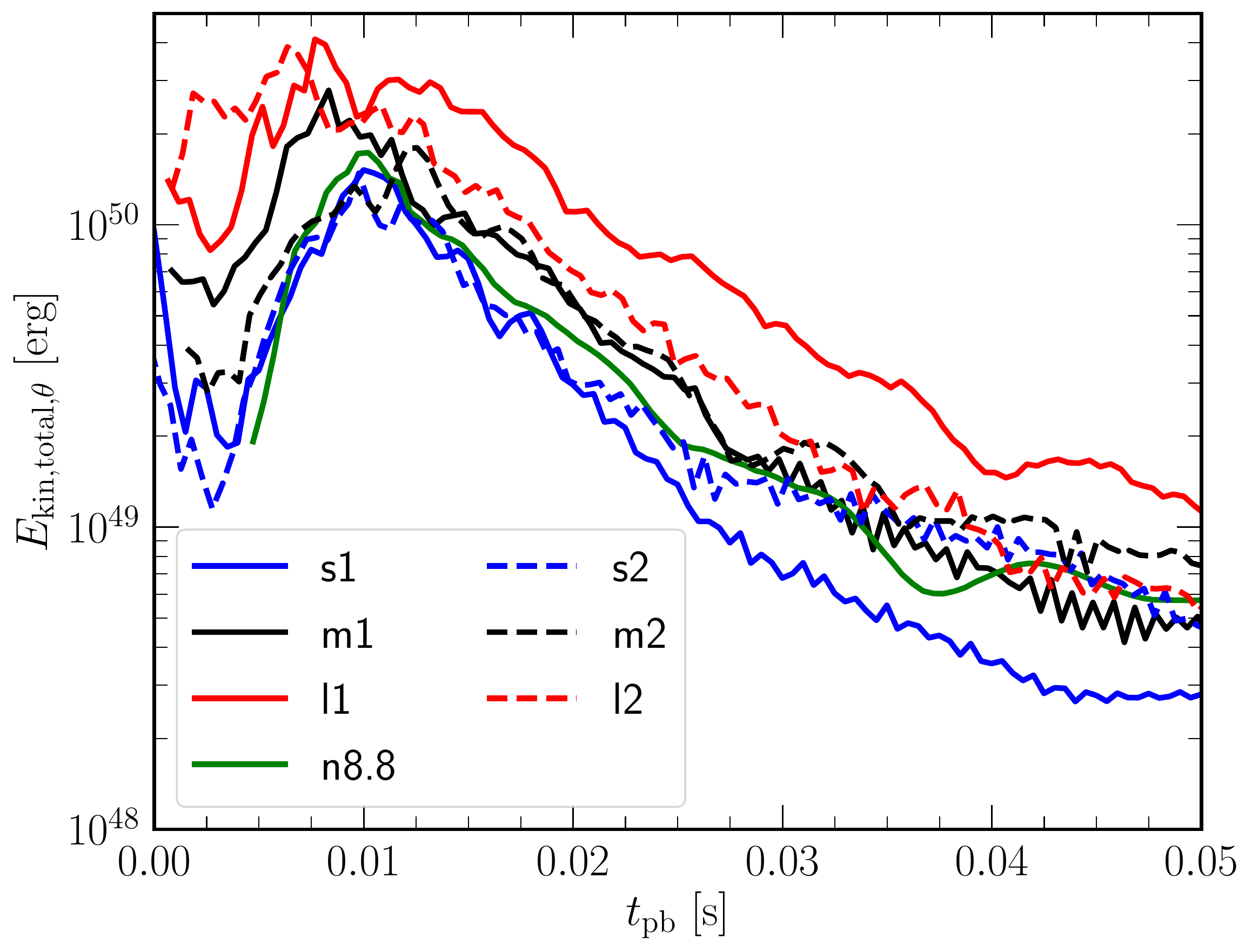}
    \includegraphics[width=0.47\textwidth]{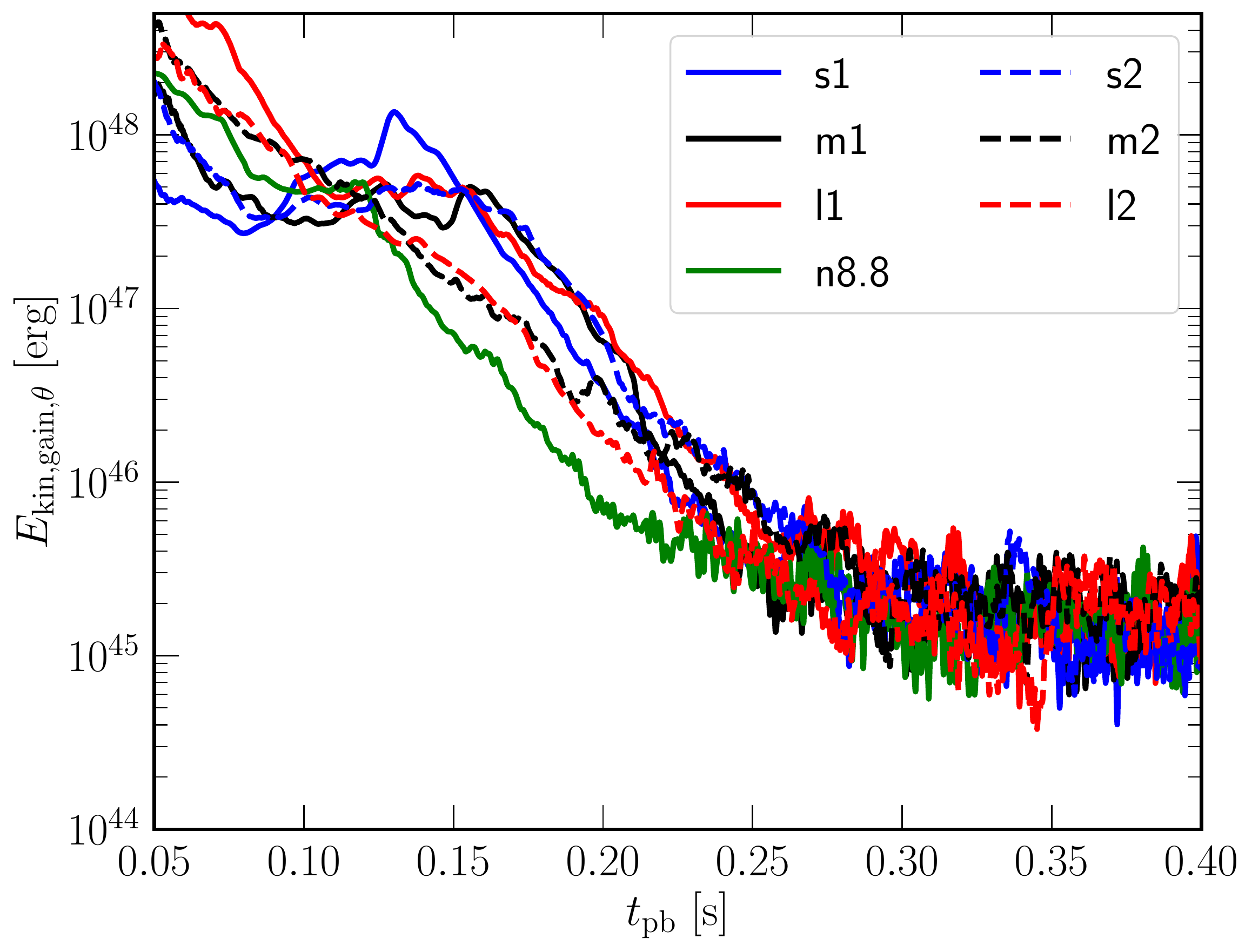}
    \caption{Left panel: Total transverse kinetic energy of the matter inside the shock as a function of the postbounce time in the 0.05~s after bounce. Right panel: Same as the left panel, but for the matter in the gain region from 0.05~s to 0.4~s after bounce.}
    \label{fig:eturb}
\end{figure*}

It is more instructive to look at the difference in the total kinetic energy of the transverse motion ($E_{\rm kin,\theta}$) in turbulent regions, which serves as a proxy for the overall turbulence. Because in the early postbounce phase the prompt convection spans from $\sim10$~km to the shock, we use $E_{\rm kin,total,\theta}$ for matter inside the shock as a measure of turbulence in the 0.05~s after bounce (left panel of Fig.~\ref{fig:eturb}). It manifests the impact of precollapse asphericities due to the oxygen flame on the prompt convection in the postshock region. Generally, a larger flame front leads to stronger turbulence in the postshock region when the aspherical perturbations pass through the shock. While models with a small flame front have the same strength of prompt convection as model n8.8, models with a large flame front acquire a larger turbulent energy by a factor of 2. In the left panel of Fig.~\ref{fig:eturb}, the slightly larger $E_{\rm kin,\theta}$ seen in the model n8.8 compared with models s1 and s2 suggests that the turbulent power seeded by the numerical perturbation of our cylindrical grid is at least comparable to that seeded by asphericities of the flame with the smallest propagated radius.

At later times when convection develops rigorously in the PNS surface region ($t_{\rm pb}\gtrsim$0.05~ms), one can separate $E_{\rm kin,\theta}$ in the gain region and the PNS convective region. In the right panel of Fig.~\ref{fig:eturb} we plot $E_{\rm kin,gain,\theta}$ for the matter in the gain region for $t_{\rm pb}=$0.05-0.4~s. All simulations with 2D progenitors still possess a larger $E_{\rm kin,\theta}$ than that with model n8.8 but there is no clear relation between $E_{\rm kin,gain,\theta}$ and initial flame size. This is likely due to a joint effect of neutrino heating and turbulence. After $t_{\rm pb}\approx0.25$~s, $E_{\rm kin,\theta}$ becomes similar in all simulations as the neutrino heating becomes weak.

\begin{figure}
    \centering
    \includegraphics[width=0.47\textwidth]{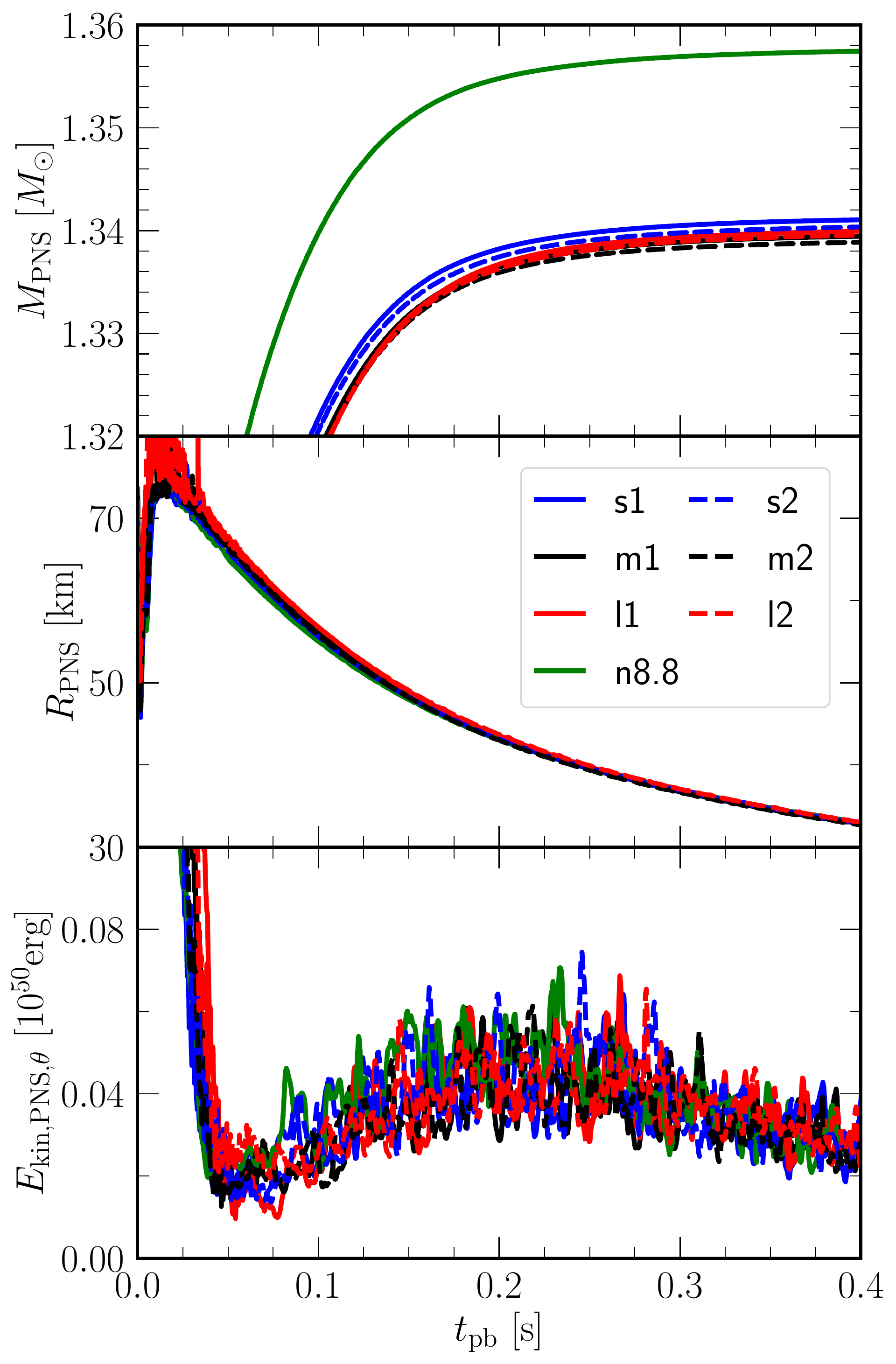}
    \caption{Same as Fig.~\ref{fig:exp}, but for the mass (top panel), radius (middle panel) and turbulent kinetic energy (bottom panel) of the protoneutron star as a function of the postbounce time in axisymmetric simulations. Note that at early postbounce time $E_{\rm kin,PNS,\theta}$ includes contribution of prompt convection.}
    \label{fig:pns}
\end{figure}

In Figure~\ref{fig:pns}, we plot the mass, radius, and transverse kinetic energy of the PNS as a function of the postbounce time. When the simulations end, the PNS baryonic mass differs by at most $\sim0.002~M_\odot$ among 2D progenitor models, which reflects the same difference in the ONeMg core mass (cf. Table~\ref{tab:prog}). This means that non-rotating ECSNe produce a NS with a generic baryonic mass of $\sim 1.34~ M_\odot$, irrespective of the uncertainty in the flame physics. The corresponding gravitational mass is $\sim1.23~M_\odot$ for a cold and $\beta$-equilibrated neutron star with the same baryonic mass and the SFHo EOS. The updated weak interaction rates used in \cite{2019ApJ...886...22Z} lead to a lower $Y_e$ and a smaller mass of the ONeMg core than that of n8.8. The latter results in a PNS baryonic (gravitational) mass of 1.357 (1.245)~$M_\odot$\footnote{We note that \cite{2017ApJ...850...43R} determined a final PNS baryonic mass of $1.30~M_\odot$ with the n8.8 progenitor. This is due to the misinterpretion of ``radius" in the data file which results in an ONeMg core mass of $1.32~M_\odot$ \citep{Radice2022}. The ``radius" in n8.8 is given at the inner edge of each mass zone.}. The PNS radius has almost no difference after $\sim$200~ms postbounce and the transverse kinetic energy shows a small difference among all simulations. The strength of PNS convection, as measured by $E_{\rm kin,PNS,\theta}$, reaches a maximum value of 4-5$\times10^{48}$ erg at 0.2~s postbounce, which is similar to that in low-mass CCSN models but about an order of magnitude smaller than more massive progenitors \citep{2020MNRAS.492.5764N,2021arXiv210609734E} due to the absence of PNS accretion. Note that vigorous PNS convection develops from $\sim0.05$~s after bounce and the early postbounce $E_{\rm kin,PNS,\theta}$ is due to the prompt convection.

\subsection{Three-dimensional simulation} \label{subsec:3D}
To explore the potential impact of the imposed axisymmetry on the core-collapse phase of ECSNe in our 2D simulations, we performed a 3D simulation in an octant with m1 as the progenitor model. The 3D simulation has a coarser resolution (see Section~\ref{subsec:sim}) than the 2D runs in Section~\ref{subsec:hydro}, so we additionally performed a 2D simulation (m1-2d-lr) with a similar resolution to disentangle the effects of resolution and dimension.

\begin{figure}
    \centering
    \includegraphics[width=0.49\textwidth]{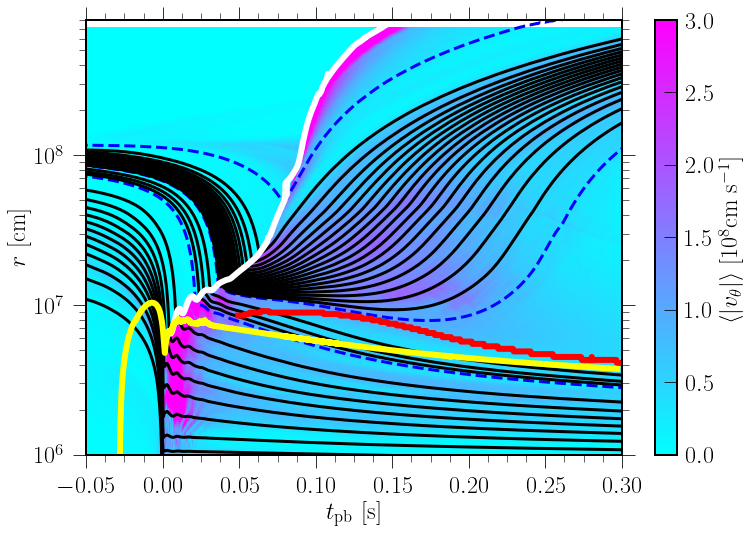}
    \caption{Same as Fig.~\ref{fig:massshell2d}, but for the magnitude of angular averaged transverse velocity $\langle |v_\theta|\rangle$ as a function of radius and the postbounce time in model m1-3d-lr. The mass resolution is same as the right plot in Fig.~\ref{fig:massshell2d} for model m1, namely 0.1~$M_\odot$, 0.01~$M_\odot$ and 0.001~$M_\odot$ below $M=$1.3, 1.34 and 1.359~$M_\odot$, which are indicated by the blue dashed lines. The white, red and yellow lines mark the mean shock radius, mean gain radius and PNS surface.}
    \label{fig:massshell3d}
\end{figure}

Fig.~\ref{fig:massshell3d} is the 3D counterpart of the right panel in Fig.~\ref{fig:massshell2d}. It is obvious that the explosion dynamics and structure of turbulent convection are similar in the 2D and 3D simulations. Compared to the 2D results, in 3D $\langle |v_\theta| \rangle$ looks smoother in the gain region during 0.1-0.2~s after bounce and the PNS convective region is more extended. The magnitude of $\langle |v_\theta| \rangle$ is comparable between 2D and 3D simulations.                              

\begin{figure}
    \centering
    \includegraphics[width=0.47\textwidth]{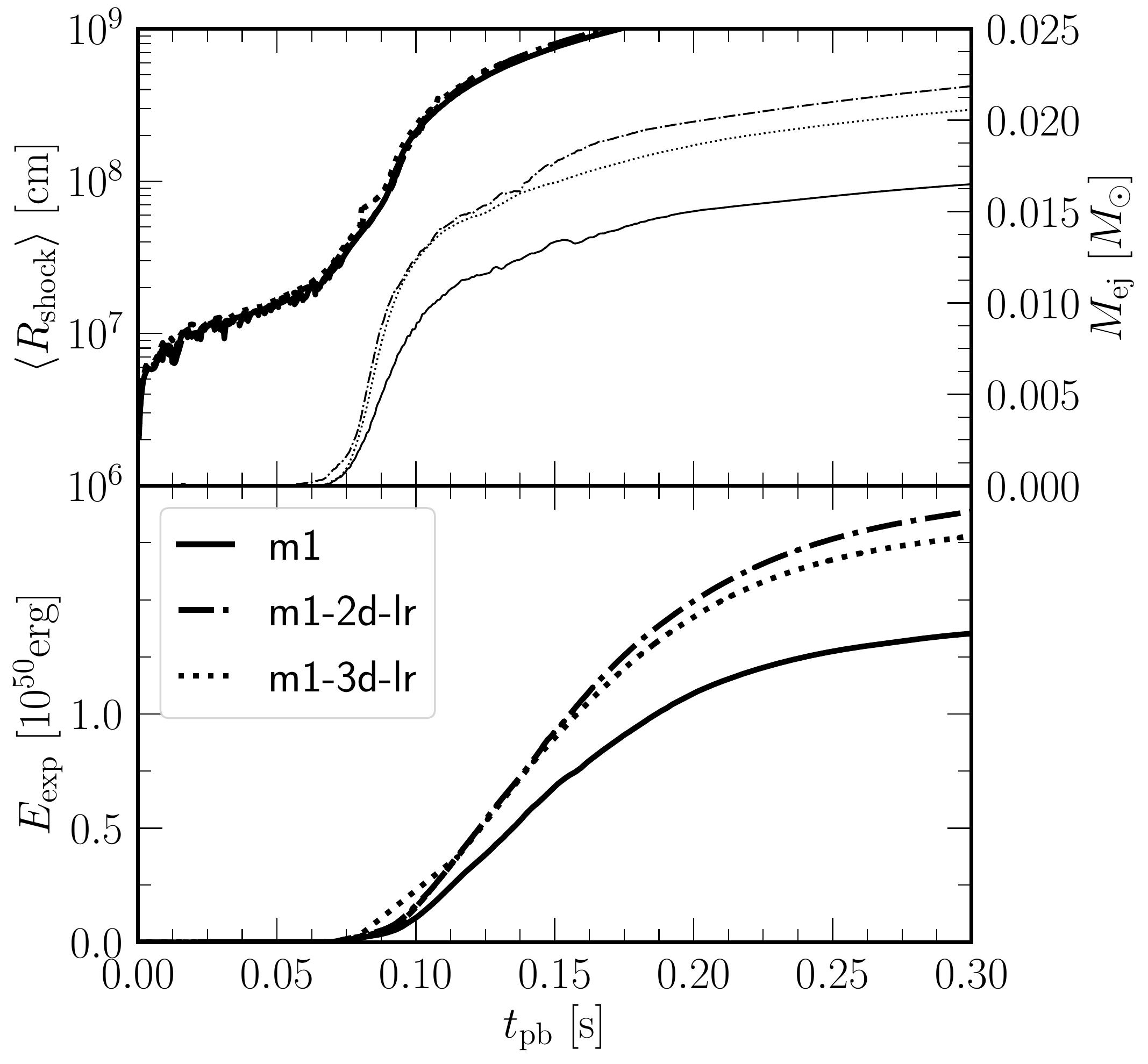}
    \caption{Same as Fig.~\ref{fig:exp}, but for the postbounce evolution of the mean shock radius, mass of ejecta and explosion energy in models m1 (solid), m1-2d-lr (dash-dotted) and m1-3d-lr (dotted). Note that the discrepancy is small between 2D and 3D simulations with a similar resolution.}
    \label{fig:exp2}
\end{figure}

\begin{figure}
    \centering
    \includegraphics[width=0.47\textwidth]{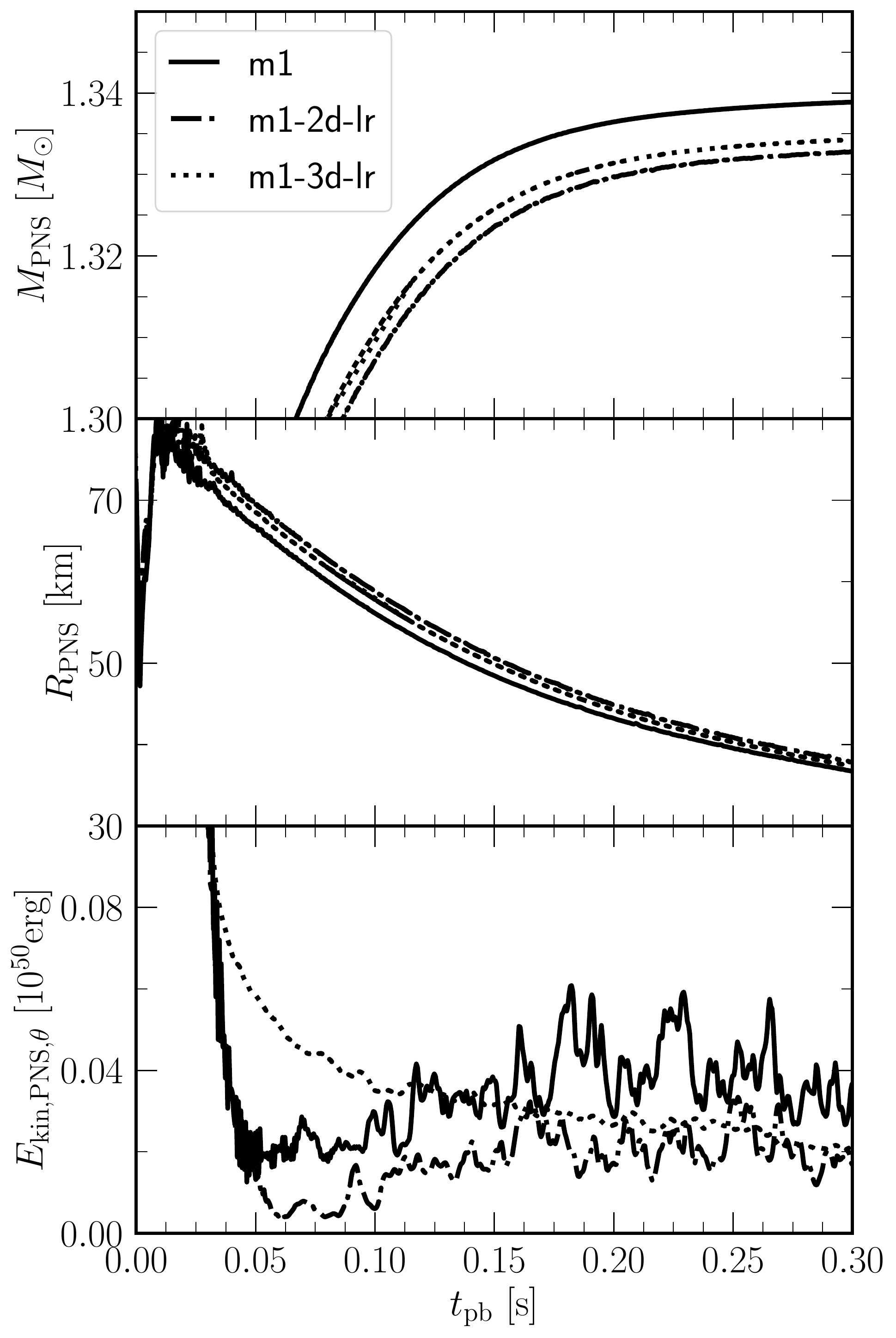}
    \caption{Same as Fig.~\ref{fig:pns}, but for the postbounce evolution of the mass, radius and transverse kinetic energy of PNS in models m1 (solid), m1-2d-lr (dash-dotted) and m1-2d-3d (dotted). Note that at early postbounce time $E_{\rm kin,PNS,\theta}$ includes contribution of prompt convection. The difference between m1-3d-lr and m1-2d-lr at $t_{\rm pb}=0.05-0.15$~s is due to the more extended convective regions in 3D (see Fig.~\ref{fig:massshell3d}) which makes it more difficult to separate the PNS and postshock convection. }
    \label{fig:pns2}
\end{figure}

As listed in Table~\ref{tab:hydro}, when the 3D simulation terminates at 300~ms after bounce, more mass (by $\sim$20\%) is ejected with a larger explosion energy (by $\sim30\%$) than the fiducial 2D simulation with the same progenitor. However, when compared to the low-resolution 2D counterpart (m1-2d-lr), the numerical results are more compatible, see Fig.~\ref{fig:exp2} for the explosion dynamics and Fig.~\ref{fig:pns2} for PNS properties. This suggests that the differences seen here are \emph{not} due to dimension but result from different resolutions.

\begin{figure}
    \centering
    \includegraphics[width=0.47\textwidth]{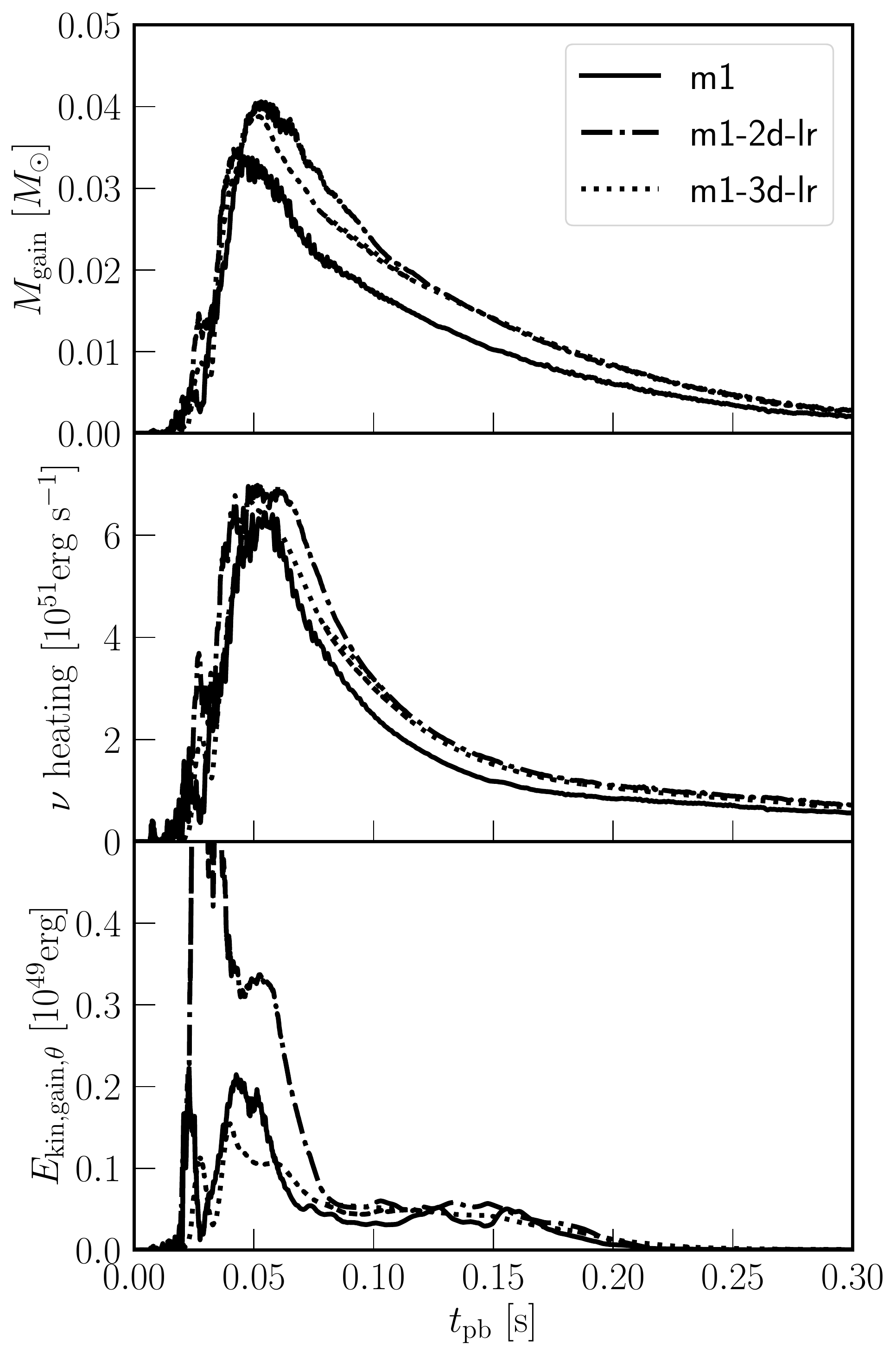}
    \caption{Mass (top panel), neutrino heating rate (middle panel) and transverse kinetic energy (bottom panel) in the gain region as a function of the postbounce time for model m1 (solid), m1-2d-lr (dash-dotted) and m1-3d-lr (dotted). Note that for the 3D simulation, the transverse kinetic energy includes both $\theta$- and $\phi$-direction motions.}
    \label{fig:2Dvs3D}
\end{figure}

To have a closer look at the difference between 2D and 3D simulations, in Fig.~\ref{fig:2Dvs3D} we compare the mass, neutrino heating rate and transverse kinetic energy $E_{\rm kin,gain,\theta}$ in the gain region for the model m1, m1-2d-lr and m1-3d-lr. Note that for the 3D simulation, $E_{\rm kin,gain,\theta}$ includes both $\theta$- and $\phi$-direction contributions. There is a visible difference between simulations with the fiducial resolution and low resolution (2D and 3D), especially after $t_{\rm pb}\approx$0.05~s. The larger mass and heating rate in the gain region are due to a less compact PNS in low-resolution runs (see Fig.~\ref{fig:pns2}). The difference in $E_{\rm kin,gain,\theta}$ can result from both different heating rates and angular resolutions ($0.6^{\circ}$ vs. $1.5^{\circ}$). 

For the low-resolution simulations m1-2d-lr and m1-3d-lr, the large discrepancy of $E_{\rm kin,gain,\theta}$ in the early postbounce phase is likely due to a different manifestation of the violent prompt convection inside the shock. Afterwards, quantitative agreement is found between the low-resolution runs in the episode when the explosion energy increases quickly ($\sim0.1-0.3$~s after bounce). Our results indicate a minor impact of 3D over 2D simulations for the collapse and explosion phases of ECSN, comparing to the 25\%-30\% difference between 2D and 1D simulations \citep[see also][]{2017ApJ...850...43R}.

\subsection{Properties of ejecta} \label{subsec:ejecta}

\begin{figure*}
    \centering
    \includegraphics[width=1.0\textwidth]{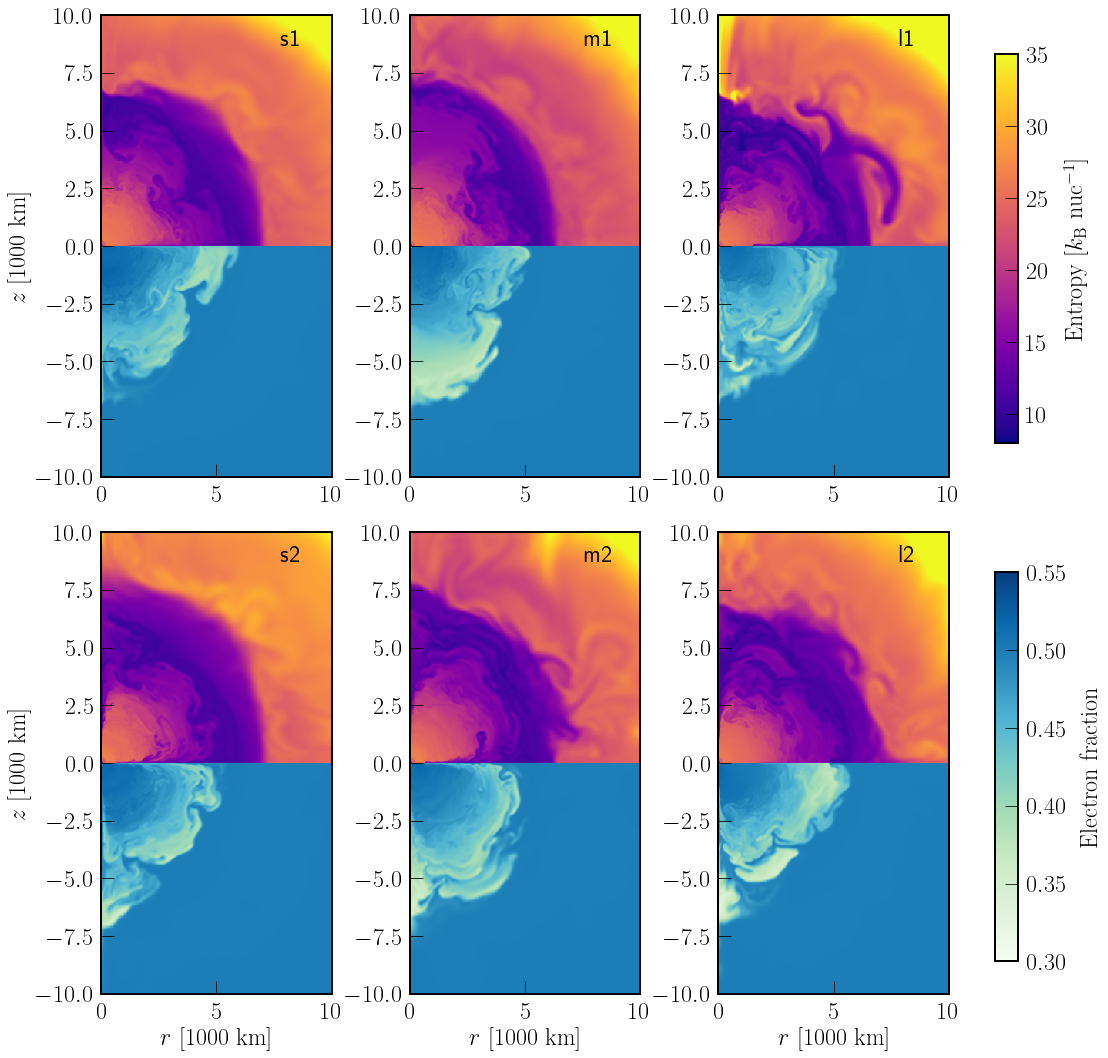}
    \caption{Slices of entropy (upper panel) and $Y_e$ (lower panel) at 300~ms postbounce in the 6 axisymmetric simulations.}
    \label{fig:pb_slice}
\end{figure*}

\begin{figure*}
    \centering
    \includegraphics[width=0.75\textwidth]{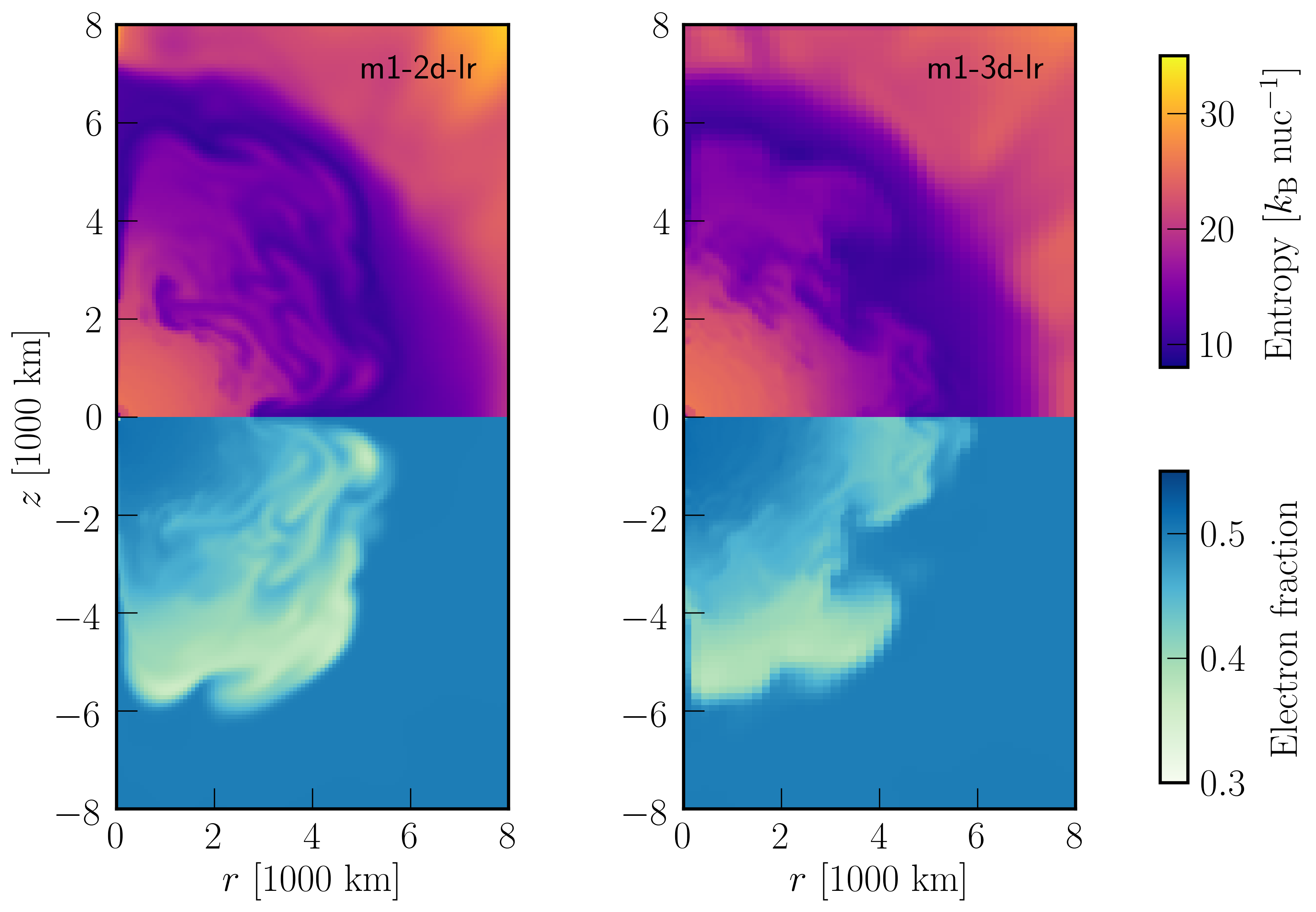}
    \caption{Similar as Fig.~\ref{fig:pb_slice}, but for the slices of entropy (upper panel) and $Y_e$ (lower panel) at 300~ms postbounce in the model m1-2d-lr and m1-3d-lr. Note that for m1-3d-lr, the $r$-axis stands for the diagonal in the $xy$-plane.}
    \label{fig:pb_slice_lr}
\end{figure*}

It has been suggested that ECSNe can be a site for $r$-process nucleosynthesis due to the early shock revival and prompt explosion \citep{1984A&A...133..175H,2003ApJ...593..968W,2007ApJ...667L.159N}. However, recent ECSN simulations with accurate modelling of neutrino transport \citep{2006A&A...450..345K,2008A&A...485..199J} have shown that it is unlikely to eject material with a low $Y_e$ ($\ll0.4$). A more detailed study showed that ECSN can produce abundant light trans-Fe elements from Zn to Zr \citep{2018ApJ...852...40W}. Here we present the properties of the ejecta in our simulations and the implications for the ECSN nucleosynthesis.

In Fig.~\ref{fig:pb_slice}, we show the 2D profiles of entropy and $Y_e$ at 300~ms after bounce in the 6 axisymmetric simulations with our fiducial resolution. The ejecta contains three components, which are the neutrino-driven PNS wind, neutrino-heated and early-shocked, from small to large radius. The early-shocked ejecta has a $Y_e$ of $\sim$0.5 and entropy from high ($\sim35~k_{\rm B}~{\rm nuc}^{-1}$) to low ($\sim10~k_{\rm B}~{\rm nuc}^{-1}$). The neutrino-heated ejecta has a $Y_e$ of $\sim$0.35-0.5 and an entropy of $\sim10$--$20~k_{\rm B}~{\rm nuc}^{-1}$. This component has inherited the aspherical structure from the neutrino-driven turbulent convection. The later ejecta in the neutrino-driven PNS wind is proton-rich with a $Y_e$ slightly above 0.5 ($\lesssim0.52$) and an entropy $\gtrsim20~k_{\rm B}~{\rm nuc}^{-1}$. There is a modest and subtle dependence of the morphology of ejecta on the employed progenitor model, especially for the neutrino-heated component. In Fig.~\ref{fig:pb_slice_lr}, we show the slices of entropy and $Y_e$ for the low-resolution runs at 300~ms after bounce. Note that for the 3D model, the $r$-axis stands for the diagonal in the $xy$-plane.

\begin{figure*}
    \centering
    \includegraphics[width=0.98\textwidth]{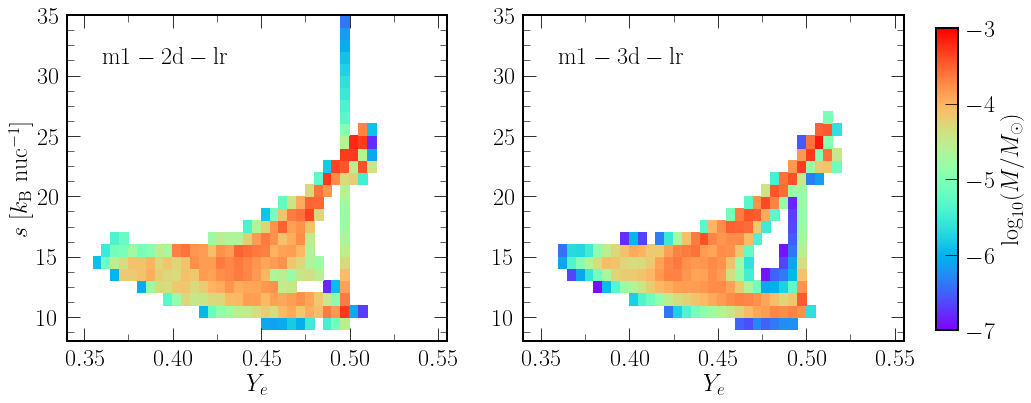}
    \caption{Mass distribution of ejecta as a function of the electron fraction ($Y_e$) and entropy per baryon ($s$) at 300~ms postbounce in models m1-2d-lr and m1-3d-lr.}
    \label{fig:ejecta_ys}
\end{figure*}

To diagnose the impact of 3D, in Fig.~\ref{fig:ejecta_ys} we compare the mass distribution of ejecta in the $Y_e$-entropy space at 300~ms after bounce in m1-2d-lr (left panel) and m1-3d-lr (right panel). The lack of high-entropy ejecta with $Y_e\approx0.5$ (vertical strike at $Y_e\approx0.5$) in m1-3d-lr is due to matter that has left the computational grid, with a small domain in 3D ($8\times10^{3}$~km) than in 2D ($1.2\times10^{4}$~km). Otherwise the mass distribution looks very similar between 2D and 3D simulations and is consistent with the slice view of the fiducial 2D runs (Fig.~\ref{fig:pb_slice}).

\begin{figure*}
    \centering
    \includegraphics[width=0.47\textwidth]{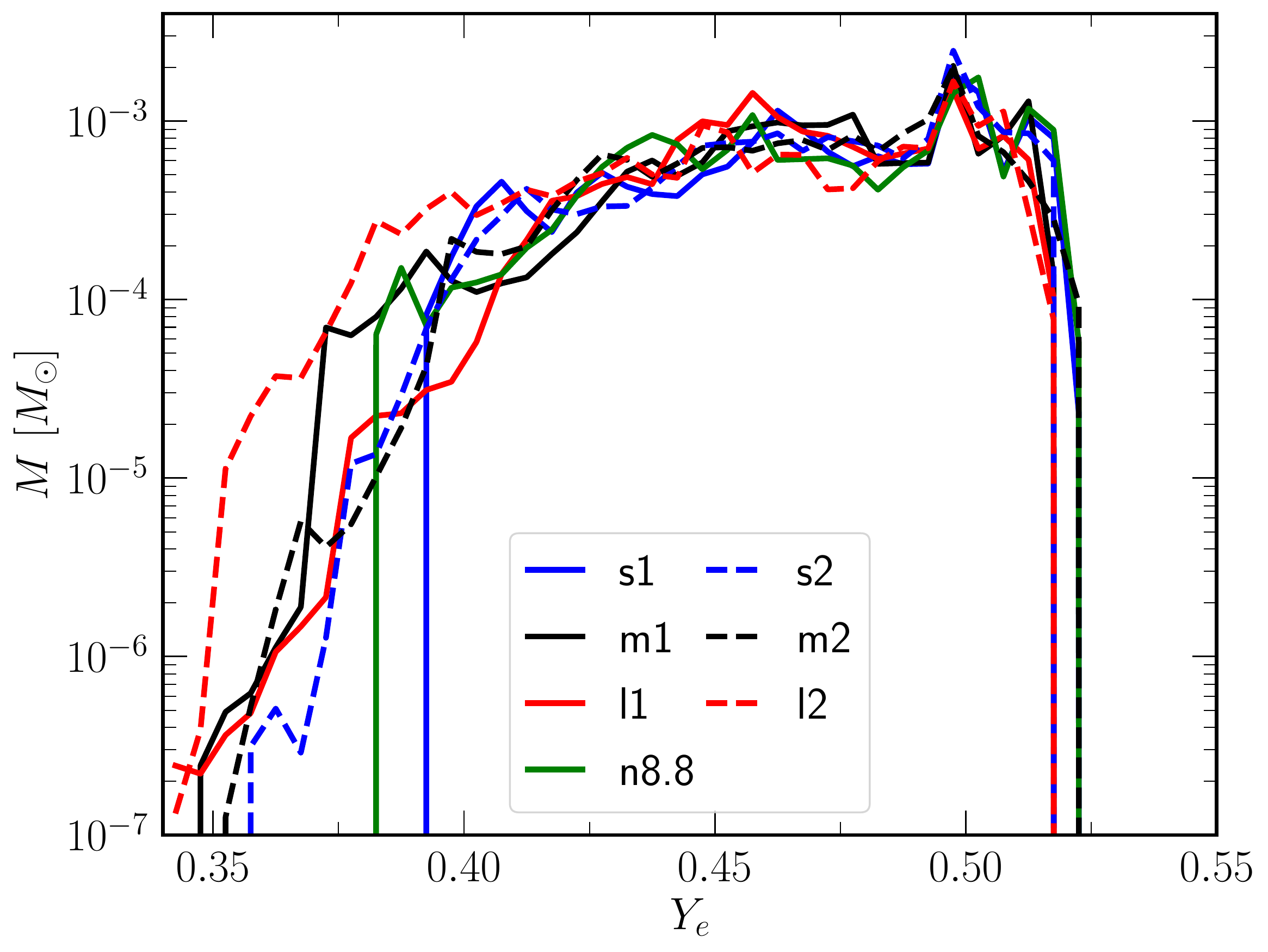}
    \includegraphics[width=0.47\textwidth]{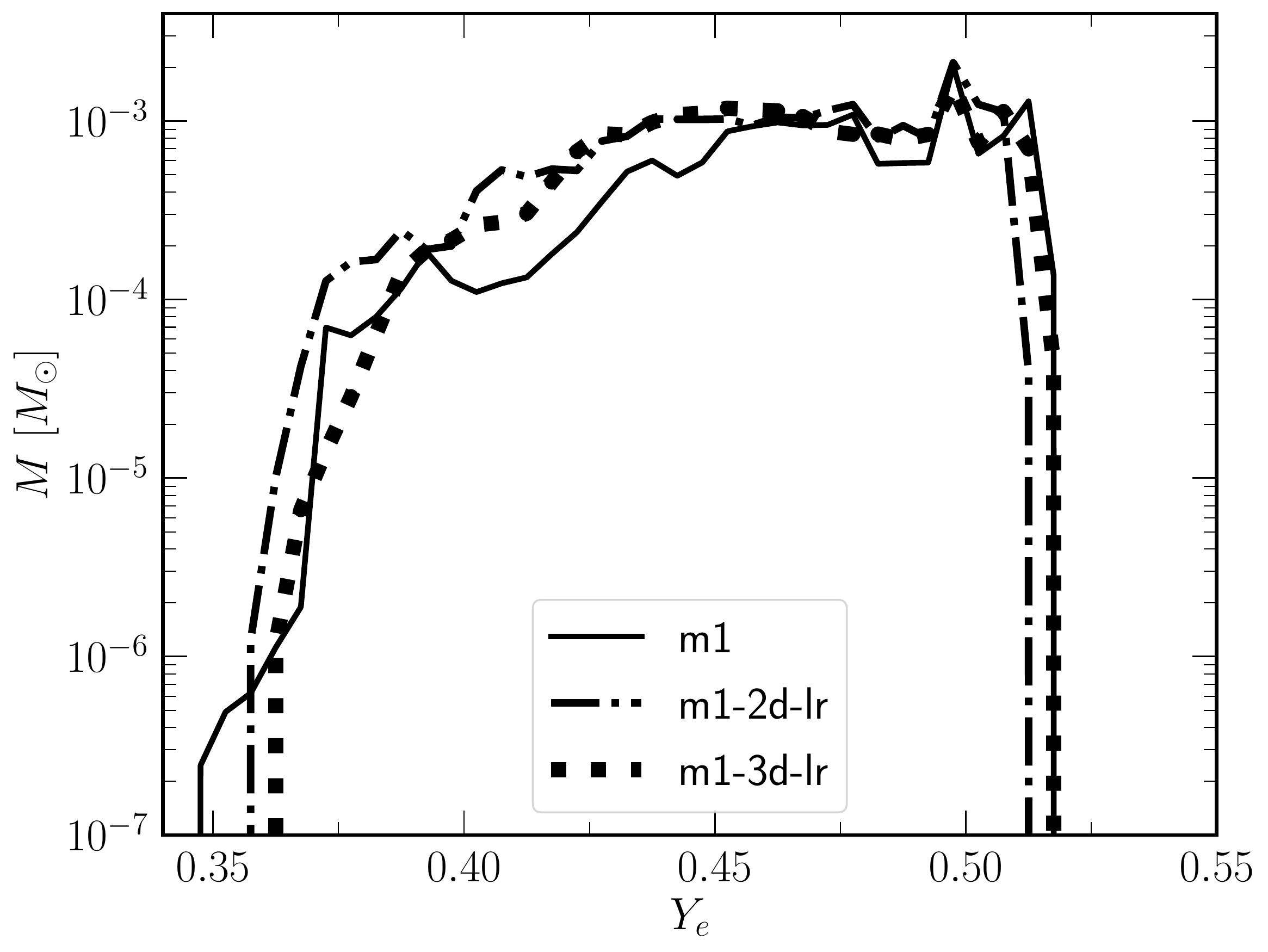}
    \caption{Mass distribution of ejecta as a function of the electron fraction ($Y_e$) in all our multidimensional ECSN simulations. The width of binning in $Y_e$ is 0.005.}
    \label{fig:ejecta_ye}
\end{figure*}

\begin{table*}
    \centering
    \begin{tabular}{ccccc}
    \hline
        Model & \multicolumn{4}{c}{{$M_{\rm ej}~(10^{-3}M_\odot)$}} \\
        & {$Y_e\le0.40$}  & {$0.4<Y_e\le0.45$}  & {$0.45<Y_e\le0.5$} & {$Y_e>0.5$} \\ \hline
         n8.8-1d  &   0  &   0   & 11.68 & 1.77  \\
            m1-1d &   0  &   0   & 11.59 & 1.46  \\ \hline
            n8.8  & 0.53 & 5.03 & 8.32 & 2.61    \\
            s1    & 0.59 & 4.18 & 9.11 & 2.40    \\
            s2    & 0.47 & 4.47 & 9.71 & 2.32    \\
            m1    & 0.76 & 4.11 & 9.34 & 2.25    \\
            m2    & 0.49 & 4.91 & 9.10 & 1.51    \\
            l1    & 0.19 & 5.19 & 9.07 & 1.54    \\
            l2    & 1.82 & 5.52 & 7.27 & 1.52    \\ \hline
            m1-2d-lr & 1.55 & 7.78 & 11.32 & 1.16   \\
            m1-3d-lr & 0.92 & 7.75 & 9.88 & 1.87 \\ \hline
    \end{tabular}
    \caption{Detailed mass distribution of ejecta as a function of $Y_e$ at 300 ms after bounce in all our simulations (cf. Table~\ref{tab:hydro}).}
    \label{tab:mej}
\end{table*}

In Fig.~\ref{fig:ejecta_ye}, we plot the mass distribution of ejecta as a function of $Y_e$ at 300~ms after bounce for all the simulations. The left panel in Fig.~\ref{fig:ejecta_ye} provides an overview for the progenitor dependence of ejecta. Progenitor models with a large $R_{\rm flame}$ (groups m and l) result in an extended tail at $Y_e\lesssim0.4$ down to $\sim$0.35. The total mass in this low-$Y_e$ tail depends also on the early shock expansion after bounce so that it shows a non-monotonic trend with respect to $R_{\rm flame}$. For model l2 which ejects the most low-$Y_e$ material, ejecta with $0.35\le Y_e<0.4$ has a mass of $\sim1.8\times10^{-3}~M_\odot$ and an entropy of 10-15~$k_{\rm B}~{\rm nuc}^{-1}$, belonging to the neutrino-heated ejecta. This may contribute to some neutron-rich nucleosynthesis and deserves further calculation with careful post-process calculation. Matter with $0.4<Y_e<0.52$ comprises the bulk ($\sim99\%$ for l1 to $\sim91\%$ for l2) of ejecta. The 3D result (m1-3d-lr) is consistent with that of m1-2d-lr (right panel of Fig.~\ref{fig:ejecta_ye}) except for a little less low-$Y_e$ component. The low-resolution simulations (both in 2D and 3D) yield a larger mass of ejecta with $0.4\lesssim Y_e\lesssim0.45$, in agreement with their more powerful neutrino heating. This again shows the minor impact of 3D on ECSN explosions. 

Overall, our results are fairly compatible with \cite{2011ApJ...726L..15W,2018ApJ...852...40W} though two differences are found. One is the low-$Y_e$ ejecta ($Y_e<0.4$) due to the 2D progenitor models of \cite{2019ApJ...886...22Z} employed in this work, which have an aspherical flame front and larger $R_{\rm flame}$ compared to model n8.8. The other is that the neutrino-driven wind has a $Y_e$ of $\lesssim0.52$ in this work while its maximum $Y_e$ reaches $\sim$0.55 in \cite{2018ApJ...852...40W}. This may be due to the different neutrino-transport schemes used in this work and \cite{2018ApJ...852...40W} that deserves further exploration.

\subsection{Discussion}
The degenerate ONeMg core of ECSN progenitors possess an outwardly propagating oxygen flame, which is not captured in our neutrino-transport simulations. According to \cite{2016A&A...593A..72J} and \cite{2020ApJ...889...34L}, the flame speed, either laminar or turbulent, is less than $\sim0.1c_{\rm s}\simeq10^{-3}c=300~$km~s$^{-1}$, where $c_{\rm s}$ ($c$) is the speed of sound (light). Therefore, before the shock breaks out of the core surface which takes $\sim0.15$~s from the start of simulations, the flame front can propagate outwards for $\sim45$~km at most. The propagated distance is small compared with both the radius of the initial flame front $R_{\rm flame}$ and the difference in $R_{\rm flame}$ among different progenitor models employed in this work. This suggests that the ignorance of the propagating oxygen flame should not affect the properties of ECSN explosion too much. 

On the other hand, the deflagration wave might transit to a detonation wave with a supersonic propagation speed, similar to that in a Type Ia supernova \citep{1991A&A...246..383K}. If such a transition would happen in an ECSN, it might form a shock that could eject the outer layer. However, the propagation of the oxygen flame through $R_{\rm flame}$ as simulated in \cite{2019ApJ...886...22Z} does not reach the condition of the deflagration-detonation transition (DDT) adopted in
\cite{2018ApJ...861..143L} and \cite{2020ApJ...889...34L}.  This is because the nuclear fuel is oxygen which is more difficult to ignite compared with carbon and also the density at the oxygen flame front is higher and less turbulent compared with the Type Ia supernova model, although the condition of the DDT is highly uncertain \citep{2017hsn..book.1185R}. We also note that when including the flame propagation in a spherically-symmetric ECSN simulation, \cite{2019ApJ...871..153T} found an enhancement of the flame speed to $\sim1000$~km~s$^{-1}$ due to neutrino-electron scattering. This new driven mechanism for flame propagation has not been confirmed elsewhere, and should be further checked in multi-dimensional neutrino-transport simulations with a proper flame-capturing scheme.

Our simulations have treated the matter outside the flame front as iron for densities above $4\times10^6$g~cm$^{-3}$ and ignored nuclear burning at lower densities. Things are flashed into NSE composition when crossing the transition density of EOS. The binding energy difference between $^{56}$Ni and O-Ne of 0.1~$M_\odot$ is $\sim1.4\times10^{50}$~erg, which is comparable to the ECSN explosion energy found here. Nevertheless, the explosion energy shows a small difference ($\sim0.1\times10^{50}$~erg) among progenitor models with a range of mass of the NSE ash (0.09 to 0.79 $M_\odot$, cf. Table~\ref{tab:prog}). It suggests that as most ($\sim98.5\%$) of the ONeMg core is quickly accreted by the PNS and remains bounded, the detailed composition has a minor effect on the explosion energy and dynamics. We leave a proper treatment for materials outside the flame front \citep[e.g.,][]{2008A&A...485..199J} for a future study. 

\section{Conclusions} \label{sec:conclu}
In this paper, we have presented 2D axisymmetric ECSN simulations using \texttt{FLASH} with progenitor model n8.8 of \cite{1984ApJ...277..791N,1987ApJ...322..206N} and six 2D progenitor models of \cite{2019ApJ...886...22Z}. We found a weak dependence of the properties of ECSN explosions on progenitor models. The diagnostic explosion energy at the termination of simulations (400~ms after bounce) is $\sim1.36$--$1.48\times10^{50}$~erg with $\sim0.017$--$0.018M_\odot$ being ejected, in the range of results reported previously in \cite{2008A&A...485..199J,2010A&A...517A..80F,2017ApJ...846..170T}. The PNS has a baryonic mass of 1.34~$M_\odot$ with the new progenitor models, slightly lighter than that with n8.8 ($\sim1.36M_\odot$). This reflects the difference in the ONeMg core mass due to updated weak-interaction rates. In addition, the extended flame front leads to more violent convective motions behind the shock, and results in a low-$Y_e$ (0.35-0.4) tail in the mass distribution of the ejecta. This low-$Y_e$ ejecta has a maximum mass of $\sim1.8\times10^{-3}~M_\odot$ and may contribute to some weak neutron-rich nucleosynthesis, which deserves further investigation with post-process calculations.

With the 3D octant simulation and its 2D counterpart with a similar resolution, we demonstrated that 3D has a small impact on the dynamical processes in ECSN, while multi-dimensional simulations result in 25\%-30\% difference in explosion properties comparing to 1D simulations. The major phenomenon observed in the 3D run is the smoother and more extended turbulent structure inside convective zones than that in 2D simulations. We attribute the small impact of 3D (comparing to the case of CCSN simulations) to the subdominant role of turbulence in the ECSN explosion. Future studies can investigate the effect of resolution in 3D ECSN simulations, perhaps in $4\pi$ geometry and with a true 3D progenitor model \citep{2016A&A...593A..72J}.  

Finally, we expect that our neglect of the flame propagation after the onset of the ONeMg core collapse will not significantly alter our conclusions. Nevertheless, rigorous modelling of the nuclear burning during the collapse and explosion is needed to yield more accurate numerical results for the explosion energy, the mass and composition of the ejecta, and the PNS mass, which are the necessary inputs for the predictions of electromagnetic signals and nucleosynthesis of ECSN.

\section*{Acknowledgements}
This work is supported by the Swedish Research Council (Project No. 2020-00452). The simulations were enabled by resources provided by the Swedish National Infrastructure for Computing (SNIC) at PDC and NSC partially funded by the Swedish Research Council through grant agreement No. 2016-07213. S.C.L. acknowledges support by NASA grants HST-946AR-15021.001-A and 80NSSC18K1017. K.N. is supported by the World Premier International Research Center Initiative (WPI), MEXT, Japan, and the Japan Society for the Promotion of Science (JSPS) KAKENHI grant JP17K05382, JP20K04024, and JP21H04499.

S.M.C. is supported by the U.S. Department of Energy, Office of Science, Office 
of Nuclear Physics, Early Career Research Program under Award Number DE-SC0015904. 
This material is based upon work supported by the U.S. Department of Energy, Office of 
Science, Office of Advanced Scientific Computing Research and Office of Nuclear Physics, 
Scientific Discovery through Advanced Computing (SciDAC) program under Award Number DE- SC0017955.
This work was supported by
the Exascale Computing Project (17-SC-20-SC), a collaborative effort
of the U.S. Department of Energy Office of Science and the National
Nuclear Security Administration.

\section*{Data Availability}
The ECSN progenitor models used in this work are publicly available at \url{https://doi.org/10.5281/zenodo.5748457}. Results of the hydrodynamic simulations will be shared on reasonable request to the corresponding author.

\bibliography{the_bib}{}
\bibliographystyle{mnras}

\bsp	
\label{lastpage}
\end{document}